\documentclass[conference]{IEEEtran}
\usepackage{amsmath}
\usepackage{todonotes}
\usepackage[T1]{fontenc}
\usepackage[noadjust]{cite}
\usepackage{bm}
\usepackage{hyperref}
\usepackage{color, colortbl}
\usepackage{xintexpr}
\usepackage{multirow}
\usepackage{etoolbox}
\usepackage[font=small]{caption}
\usepackage{paralist}
\usepackage{algorithm}
\usepackage[ruled,vlined,algo2e]{algorithm2e}
\usepackage[subtle, bibbreaks=normal, paragraphs=tight, floats=tight, mathspacing=tight, wordspacing=normal, tracking=normal ]{savetrees}
\makeatletter
\patchcmd{\@makecaption}
  {\scshape}
  {}
  {}
  {}
\makeatother

\SetKwInput{KwGiven}{Given}
\SetKwInput{KwFind}{Find}
\SetKwInput{KwWhere}{Where}
\SetKwInput{KwMin}{Minimizing}

\newcommand{\lfm}{\href{http://www.last.fm}{last.fm}~}

\newcommand{\figwidth}{.92}
\newcommand{\subfigwidth}{.45}
\newcommand{\boxscale}{.92}
\newcommand{\bannerwidth}{1}
\newcommand{\topk}{10}
\newcommand{\twotopk}{{\xinttheexpr 2*\topk{}\relax}}
\DeclareMathOperator*{\argmin}{argmin}

\definecolor{LightCyan}{rgb}{0.88,1,1}
\definecolor{Gray}{gray}{0.8}
\begin{document}

\title{Network Model Selection for Task-Focused Attributed Network Inference}

\author{\IEEEauthorblockN{Ivan Brugere}
\IEEEauthorblockA{University of Illinois at Chicago \\
ibruge2@uic.edu}
\and
\IEEEauthorblockN{Chris Kanich}
\IEEEauthorblockA{University of Illinois at Chicago\\
ckanich@uic.edu}
\and
\IEEEauthorblockN{Tanya Y. Berger-Wolf}
\IEEEauthorblockA{University of Illinois at Chicago\\
tanyabw@uic.edu}}



\maketitle

\begin{abstract}

Networks are models representing relationships between entities. Often these relationships are explicitly given, or we must learn a representation which generalizes and predicts observed behavior in underlying individual data (e.g. attributes or labels). Whether given or inferred, choosing the best representation affects subsequent tasks and questions on the network. This work focuses on model selection to evaluate network representations from data, focusing on fundamental predictive tasks on networks. We present a modular methodology using general, interpretable network models, task neighborhood functions found across domains, and several criteria for robust model selection. We demonstrate our methodology on three online user activity datasets and show that network model selection for the \textit{appropriate network task} vs. an alternate task increases performance by an \textit{order of magnitude} in our experiments.

\end{abstract}
\IEEEpeerreviewmaketitle

\section{Introduction}

Networks are models representing relationships between entities: individuals, genes, documents and media, language, products etc. We often assume the expressed or inferred edge relationships of the network are correlated with the underlying behavior of individuals or entities, reflected in their actions or preferences over time. 

On some problems, the correspondence between observed behavior and network structure is well-established. For example, simple link-prediction heuristics in social networks tend to perform well because they correspond to the social processes for how networks grow \cite{Adamic2003}. However, content sharing in online social networks shows that `weak' ties among friends account for much of the influence on users \cite{Bakshy:2012:RSN:2187836.2187907}. For these different tasks, the same friendship network ``as-is'' is a relatively better \emph{model} for predicting new links than predicting content sharing. Learning an \textit{alternative} network representation for content sharing better predicts this behavior and is more informative of the relevant relationships for one task against another.

Why should we learn a network representation at all? First, a \textit{good} network for a particular predictive task will perform better than methods over aggregate populations. Otherwise, edge relationships are not informative with respect to our purpose. Evaluating network models against population methods measures whether there is a network effect at all within the underlying data. In addition, network edges are \textit{interpretable} and suitable for descriptive analysis and further hypothesis generation. Finally, a good network model \textit{generalizes} several behaviors of entities observed on the network, and we can evaluate this robustness under a shared representation. For these reasons, we should learn a network representation if we can evaluate `which' network is useful and whether there is the presence of any network on the underlying data. 

Much of the previous work focuses on method development for better predictive accuracy on a \textit{given network}. However, predictive method development on networks treats the underlying network representation as independent of the novel task method, rather than coupled to the \textit{network representation} of underlying data. Whether network structure is given or inferred, choosing the best representation affects subsequent tasks and questions on the network. How do we measure the effect of these network modeling choices on common and general classes of predictive tasks (e.g. link prediction, label prediction, influence maximization)? 

Our work fills this methodological gap, coupling general network inference models from data with common predictive task methods in a network model selection framework. This framework evaluates both the definition of an edge, and the `neighborhood' function on the network most suitable for evaluating various tasks. 

\section{Related Work}

Our work is primarily related to research in relational machine learning, and network structure inference.

Relational learning in networks uses correlations between network structure, attribute distributions, and label distributions to build network-constrained predictive methods for several types of tasks. Our work uses two fundamental relational learning tasks, link prediction and collective classification to evaluate network models inferred from data.

The link prediction task \cite{Hasan2011} predicts edges from `local' information in the network. This is done by simple ranking, on information such as comparisons over common neighbors \cite{Adamic2003}, higher-order measures such as path similarity \cite{Liben-Nowell2007}, or learning edge/non-edge classification as a supervised task on extracted edge features \cite{al2006link, Gong:2014:JLP:2611448.2594455}.

The collective classification task \cite{london2014collective, sen:aimag08} learns relationships between local neighborhood structure of labels and/or attributes \cite{McDowell:2013:LAR:2505515.2505628} to predict unknown labels in the network. This problem has been extended to joint prediction of edges \cite{4476695}, and higher-order joint prediction \cite{namata:tkdd15}. Our work does not extend these methods. Where suitable, more sophisticated predictive task methods and higher-order joint methods may be used for evaluating network models and model selection.

Network structure inference \cite{2016arXiv161000782B, Kolaczyk2014} is a broad area of research aimed at transforming data on individuals or entities into a network representation which can leverage methods such as relational machine learning. Previous work spans numerous domains including bioinformatics \cite{ZhangHorvath2005}, neuroscience \cite{Sporns2014}, and recommender systems \cite{McAuley:2015:INS:2783258.2783381}. Much of this work has domain-driven network model evaluation and lacks a general methodology for transforming data to useful network representations.

Models for network inference are generally either parametric, or similarity-based. Parametric models make assumptions of the underlying data distribution and learn an edge structure which best explains the underlying data. For example, previous work has modeled the `arrival time' of information in a content network with unknown edges, where rates of transmission are the learned model \cite{Gomez-Rodriguez:2012:IND:2086737.2086741, Myers2010}. 

Several generative models can sample networks with correlations between attributes, labels, and network structure, and can estimate model parameters from data. These include the Attributed Graph Model (AGM) \cite{Pfeiffer:2014:AGM:2566486.2567993}, Multiplicative Attribute Graph model (MAG) \cite{Kim2012}, and Exponential Random Graph Model (ERGM) \cite{Robins2007}. However, these models are typically not estimated against a task of interest, so while it may fit our modeling assumption, it may not model the subsequent task; our proposed methodology straightforwardly accepts any of these models estimated from data.

Similarity-based methods tend to be ad-hoc, incorporating domain knowledge to set network model parameters. Recent work on task-focused network inference evaluates inferred network models according to their ability to perform a set of tasks \cite{DeChoudhury:2010:IRS:1772690.1772722}. These methods often have a high sensitivity to threshold/parameter choice, and added complexity of interactions between network representations and task models. Our work identifies these sensitivities, and yields robust model selection over several stability criteria.

\section{Contributions}
We present a general, modular methodology for model selection in task-focused network inference. Our work:
\begin{compactitem}
    \item identifies constituent components of  the common network inference workflow, including network models, tasks, task neighborhood functions, and measures for evaluating networks inferred from data,  
    \item uses fundamental, interpretable network models relevant in many application domains, and fundamental tasks and task \textit{localities} measuring different aspects of network-attribute correlations,
    \item presents significance, stability, and null-model testing for task-focused network model selection on three online user activity datasets.
\end{compactitem}

Our work focuses on process and methodology; the development of novel network models and task methods are complimentary but distinct from this work. Our work demonstrates that network representation learning is a crucial step for evaluating predictive methods on networks.

\section{Methods}

\subsection{Task-Focused Network Inference and Evaluation}

\begin{figure*}[ht]
\centering
  \includegraphics[width=\bannerwidth\textwidth]{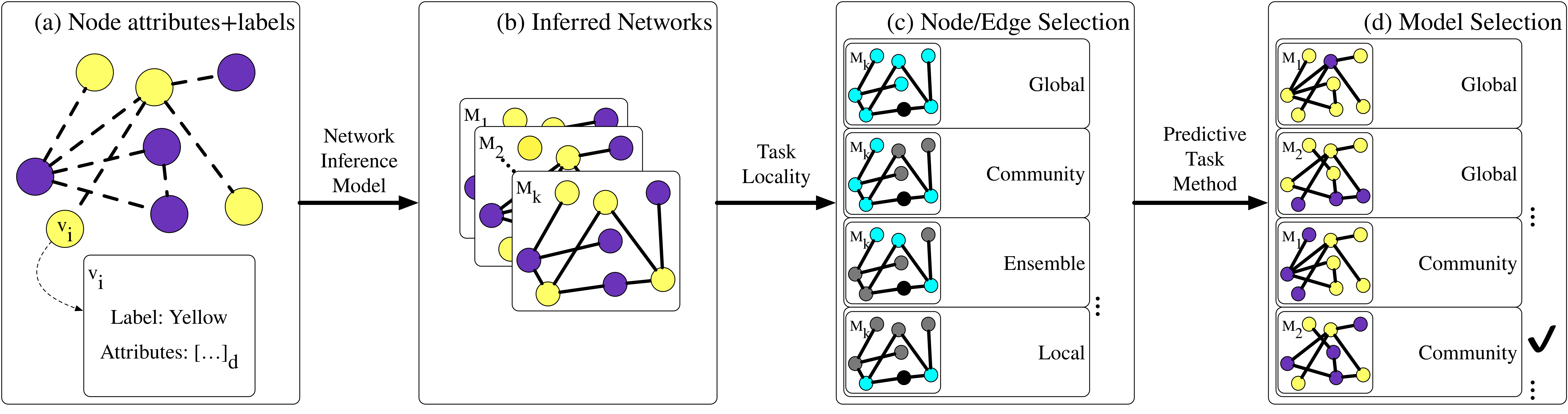}
  \caption{A schematic of the model selection methodology. (a) Individual node attributes and labels are provided as input, an edge-set may be provided (dashed edges). (b) A collection of networks are inferred from functions on attributes and labels, outputting an edge-set. (c) For each network model, we build a classification method on attributes and labels for each node of interest (black) constraining node-attribute and node-label available to the method (cyan) according to constraints from the network structure. These `global', `meso', and `local' methods produce prediction output for some task (e.g. link prediction, collective classification). (d) we select a network representation subject to performance for the given task.} 
  \label{fig:overview}
\end{figure*}

Model selection for task-focused network inference learns a network model and associated parameters which perform a task or set of tasks with high precision. We learn joint relationships between node attribute vectors and some target label (e.g. node behavior) of interest.

Figure \ref{fig:overview} presents a high-level schematic of our methodology and the constituent components of task-oriented network inference problems. First, (a) our data is a collection of attributes and labels associated with discrete entities. Edges are optionally provided to our methodology, indicated by dashed lines. Edges may be entirely missing or one or several edge-sets. In all these cases we evaluate each edge-set as one of many possible network `models.' Attributes are any data associated with nodes (and/or edges) in the network. These are often very high dimensional, for example user activity logs and post content in social networks, gene expression profiles in gene regulatory networks, or full document text and other metadata in document, video or audio content networks.

Labels are a notational convenience denoting a specific attribute of predictive interest. Labels are typically low-cardinality fields (e.g. boolean) appropriate for supervised classification methods. Our methodology accepts (and prefers) \textit{multi}-labeled networks. This simply indicates multiple fields of predictive interest. For example, we induce labels as an indicator for whether a user is a listener or a viewer of `this' music or movie genre, given by historical activity data over many such genres. Multiple sets of these labels more robustly evaluate whether attribute relationships in the fixed underlying network model generalize to these many `behaviors.' 

Second, in Figure \ref{fig:overview} (b) we apply several fundamental network inference models. These are modular and can be specified according to the domain. These models are either parametric statistical methods, or functions on similarity spaces, both of which take attributes and/or labels as input and produce an inferred edge-set. Note in \ref{fig:overview} (b) that the stacked edge-sets vary. 

Third, in Figure  \ref{fig:overview} (c) for every node of interest (black) for some predictive task, we select sets of nodes (cyan, grey nodes excluded) and their associated attributes and labels according to various \textit{task locality functions} (i.e. varying `neighborhood' or network query functions). These functions include globally selecting the population of all nodes, sampling nodes from within `this' node's community by the output of some structural community method, an ensemble of a small number of nodes according to attribute or network ranking (e.g. degree, centrality), or local sampling by simple network adjacency or some other ordered traversal. 

Once selected, the associated attributes and labels of the cyan nodes yield a supervised classification problem predicting label values from attribute vectors, using a predictive task model of interest. Several fundamental prediction problems fit within this supervised setting. For example, link prediction can use positive (edge) and negative (non-edge) label instances rather than learning classifiers on the input labelsets. 

We evaluate network models (b) over the collection of its task methods and varying localities (c) to produce Figure \ref{fig:overview} (d), some predictive output (e.g. label prediction, shown here). The model and locality producing the best performance is selected. Finally, we further evaluate the stability and significance of our selected model over the collection of possible models. 

Below, we infer networks to identify listeners/reviewers of music genres, film genres, and beer types in three user activity datasets: Last.fm, MovieLens and BeerAdvocate, respectively.





\begin{algorithm2e}
  \SetAlgorithmName{Problem}
  \\
  \\
  \KwGiven{Node Attribute-set $A$, Node Label-sets $L^*$,\\ 
  Network model-set $\mathcal{M}_j \in M$ where $\mathcal{M}_j(A,L) \rightarrow E_j'$,\\  
  Task-set $ \mathcal{C}_k \in C$ where $ \mathcal{C}_k(E_j',A,L^*) \rightarrow P_{kj}'$}
  \KwFind{Network edge-set $E_j'$}
  \KwWhere{$\argmin_j \mathcal{L}(P, P_{kj}')$ on validation $P$}
  \caption{Task-Focused Network Inference Model Selection}
  \label{p:networkinference}
\end{algorithm2e}

Problem \ref{p:networkinference} gives a concise specification of our model selection framework, including inputs and outputs. Given individual node-attribute vectors $\vec{a}_i \in A$, where `$i$' corresponds to node $v_i$ in node-set $V$, and a collection of node-labelsets $L \in L^*$, where $l_i \in L$ the label of node $v_i$ in a single labelset $L$, our general task-focused network inference framework evaluates a set of possible network models $M =\{\mathcal{M}_1...\mathcal{M}_n\}$  where each $\mathcal{M}_j: \mathcal{M}_j(A,L) \rightarrow E_j'$ produces a network edge-set $E_j'$.\footnote{Notation: capital letters denote sets, lowercase letters denote instances and indices. Primes indicate \emph{predicted} and inferred entities. Greek letters denote method parameters. Script characters (e.g. $\mathcal{C}()$) denotes functions, sometimes with specified parameters.} 

These instantiated edge-sets are evaluated on a set of inference task methods $C = \{\mathcal{C}_1...\mathcal{C}_m\}$ where each $\mathcal{C}_k$ produces $P_k'$, a collection of predicted edges, attributes, or labels  depending on context: $\mathcal{C}_k(E_j',A,L^*) \rightarrow P_{kj}'$. We evaluate a $P'$ under loss $\mathcal{L}(P, P')$, where $P$ is the validation or evaluation data. We select $\mathcal{M}_{\mathrm{select}} = \argmin_j \mathcal{L}(P, P_{kj}')$ based on performance over $C$ and/or $L^*$. Finally, we evaluate generalized performance of this model choice.

Our methodology can be instantiated under many different network model-sets $M$, task methods $C$, and loss functions $\mathcal{L}()$. We present several fundamental, \textit{interpretable} modeling choices common to many domains. All of these identified choices are modular and can be defined according to the needs of the application. For clarity, we refer to the network representation as a \textit{model}, and the subsequent task as a method, because we focus on \textit{model selection} for the former.

\subsection{Network models}

Network models are primarily parametric under some model assumption, or non-parametric under a similarity space. We focus on the latter to avoid assumptions about joint attribute-label distributions. 

We present two fundamental similarity-based network models applicable in most domains: the k-nearest neighbor (KNN) and thresholded (TH) networks. Given a similarity measure $\mathcal{S}(\vec{a}_i, \vec{a}_j) \rightarrow s_{ij}$ and a target edge count $\lambda$, this measure is used to produces a pairwise attribute similarity space. We then select node-pairs according to:


\begin{compactitem}
	\item $k$-nearest neighbor $\mathcal{M}_{\mathrm{KNN}}(A, \mathcal{S}, \lambda)$: for a fixed $v_i$, select the top $\lfloor\frac{\lambda}{|V|}\rfloor$ most similar $\mathcal{S}(\vec{a}_i, \{A \setminus \vec{a}_i\})$. In directed networks, this produces a network which is $k$-regular in out-degree, with $k=\lfloor\frac{\lambda}{|V|}\rfloor$.
	\item Threshold $\mathcal{M}_{\mathrm{TH}}(A, \mathcal{S}, \lambda)$: for \emph{any} pair $(i,j)$, select the top $\lambda$ most similar $\mathcal{S}(\vec{a}_i, \vec{a}_j)$
\end{compactitem}

For a fixed number of edges, the varying performance of these network models measures the extent that absolute similarity, or an equal allocation of modeling per node produces better predictive performance.

\subsubsection{Node attribute similarity measures}

When constructing a network model on node attribute and label data, we compare pairwise attribute vectors by some similarity and criteria to determine an edge. These measures may vary greatly according to the nature of underlying data. Issues of projections, kernels, and efficient calculation of the similarity space are complimentary to our work, but not our focus.

Our example applications focus on distributions of item counts: artist plays, movie ratings, beer ratings, where each attribute dimension is a unique item, and the value is its associated rating, play-count etc. Therefore we measure the simple `intersection' of these vectors. 

Given a pair of attribute vectors $(\vec{a}_i, \vec{a}_j)$ with non-negative elements, intersection $\mathcal{S}_{INT}()$ and normalized intersection $\mathcal{S}_{INT-N}()$ are given by:
\begin{equation}
    \begin{aligned}
	\mathcal{S}_{INT}(\vec{a}_i, \vec{a}_j) = \sum\nolimits_d \mathrm{min}(a_{id}, a_{jd}) \\
	\mathcal{S}_{INT-N}(\vec{a}_i, \vec{a}_j) = \frac{\sum\nolimits_d \mathrm{min}(a_{id}, a_{jd})}{\sum\nolimits_d \mathrm{max}(a_{id}, a_{jd})} 	
	\end{aligned}	
\end{equation}

These different similarity functions measure the extent that \textit{absolute} or \textit{relative} common elements produce a better network model for the task method. Absolute similarity favors more active nodes which are highly ranked with other active nodes before low activity node-pairs. For example, in our \lfm application below, this corresponds to testing whether the total count of artist plays between two users is more predictive than the \textit{fraction} of common artist plays.

\subsubsection{Explicit network models}

Our methodology accepts observed, explicit edge-sets (e.g. a given social network) not constructed as a function of node attributes and labels. This allows us to test given networks as fixed models on attributes and labels, for our tasks of interest. 

\subsubsection{Network density}

We evaluate network models at varying density, which has previously been the primary focus of networks inferred on similarity spaces \cite{Myers2010, DeChoudhury:2010:IRS:1772690.1772722}. When evaluating network models against a given explicit network, we fix density as a factor of the explicit network density (e.g. $0.75\times \mathrm{d}(E)$). Otherwise, we explore sparse network settings. We don't impose any density penalty, allowing the framework to equally select according to predictive performance.

\subsection{Tasks for Evaluating Network Models}

We evaluate network models on two fundamental network tasks: collective classification and link prediction. 

\subsubsection{Collective classification (CC)} The collective classification problem learns relationships between network edge structure and attributes and/or labels to predict label values \cite{sen:aimag08, 4476695}. This task is often used to infer unknown labels on the network from `local' discriminative relationships in attributes, e.g. labeling political affiliations, brand or media preferences.

Given an edge-set $E'$, a neighborhood function $\mathcal{N}(E',i)$, and node-attribute set $A$, the collective classification method $\mathcal{C}_{\mathrm{CC}}(A_{\mathcal{N}(E', i)}, L_{\mathcal{N}(E',i)}, \vec{a}_i) \rightarrow l_i'$ trains on attributes and/or labels in the neighborhood of $v_i$ to predict a label from attribute \emph{test} vector $\vec{a}_i$. 

We use this task to learn network-attribute-label correlations to identify \textit{positive} instances of rare-class labels. As an oracle, we provide the method only these positive instances because we want to learn the true rather than null association of the label behavior. Learning positive instances over many such labelsets allows us to robustly evaluate learning on the network model under such sparse labels.

\subsubsection{Link prediction (LP)}

The link prediction problem \cite{Liben-Nowell2007} learns a method for the appearance of edges from one edge-set to another. Link prediction methods can incorporate attribute and/or label data, or using simple structural ranking \cite{Adamic2003}.

Given a training edge-set $E'$ induced from an above network model, a neighborhood function $\mathcal{N}(E', i)$, and a node-attribute set $A$, we learn edge/non-edge relationships in the neighborhood as a binary classification problem on  edge/non-edge attribute vectors, $\vec{a}_{jk}$ where  $(j,k) \in \mathcal{N}(E', i)$. On input test attributes $\vec{a}_{jk}$, the model produces an edge/non-edge label: $\mathcal{C}_{\mathrm{LP}}(A_{\mathcal{N}(E', i)}, \vec{a}_{jk}) \rightarrow l_{jk}'$. For our applications, the simple node attribute intersection is suitable to construct edge/non-edge attributes: $\vec{a}_{jk} = \underset{d=1...|\vec{a}_j|}{\mathrm{min}}(a_{jd}, a_{kd})$. 

\subsection{Task Locality}
\label{subsec:task_locality}
Both of the above tasks are classifiers that take attributes and labels as input, provided by the neighborhood function $\mathcal{N}(E', i)$. However, this neighborhood need not be defined as network adjacency. We redefine the neighborhood function to provide each task method with node/edge attributes selected at varying locality from the test node. These varying localities give interpretable feedback to which level of abstraction the network model best performs the task of interest. We propose four general localities:

\subsubsection{Local}

For CC, local methods use simple network \textit{adjacency} of a node $v_i$. For LP, local methods use the \textit{egonet} of a node $v_i$. This is defined as an edge-set from nodes adjacent to $v_i$, plus the edges of these adjacent nodes. Non-edges are given by the compliment of the egonet edge-set. We also evaluate a breadth-first-search neighborhood (BFS) on each model, collecting $k$ (=200) nodes encountered in the search order.

\subsubsection{Community}

We calculate structural community labels for each network using the Louvain method \cite{ICT4DBibliography2429}. We sample nodes and edges/non-edges from the induced subgraph of the nodes in the `test' node's community. 

\subsubsection{Ensemble}

We select $k$ (=30) nodes according to some fixed ordering. These nodes become a collection of locally-trained task methods. For each test node, we select the KNN (=3) ensemble nodes according to the similarity measure of the current network model, and take a majority of their prediction.

We use decreasing node-order on (1) Degree, (2) Sum of attributes (i.e. most `active' nodes), (3) Unique attribute count (i.e. most diverse nodes), and (4) Random order (one fixed sample per ensemble).

Ensembles provide `exemplar' nodes expected to have more robust local task methods than those trained on the test node's neighborhood, on account of their ordering criteria. These nodes are expected to be suitably similar to the test node, on account of the KNN selection from the collection. 
 
\subsubsection{Global}

We sample a fixed set of nodes or edges/non-edges without locality constraint. This measures whether the network model is informative at all, compared to a single global classification method. This measures the extent that a task is better represented by aggregate population methods than encoding local relationships. Although models with narrower locality are trained on less training data, they can learn local heterogeneity of local attribute relationships which may confuse an aggregated model. 

\subsection{Classification Methods}

For all different localities, both of our tasks each reduce to a supervised classification problem. For the underlying classification method, we use standard linear support vector machines (SVM), and random forests (RF). These are chosen due to speed and suitability in sparse data. Absolute predictive accuracy is not the primary goal of our work; these methods need only to produces consistent network model ranking over many model configurations. 

\subsection{Network Model Configurations}
We define a network model configuration as a combination of all specified modeling choices: network model, similarity measure, density, task locality, and task method. Each network model configuration can be evaluated independently, meaning we can easily scale to hundreds of configurations on small and medium-sized networks.

\section{Datasets}

 \begin{table*}[t]
 	\centering
	\begin{tabular}{|c|c|c|c|c|c|}
		\hline
		Dataset & |V| & |A| & Median/90\% Unique Attributes & Median/90\% Positive Labels & Labelsets\\\hline
		Last.fm 20K \cite{mlg2017_13} & 19,990 &  1,243,483,909 artist plays & 578/1713 artists & 3/11 music genres & 60 genres\\
		MovieLens \cite{Harper:2015:MDH:2866565.2827872} & 138,493 & 20,000,263 movie ratings & 81/407 ratings & 7/53 movie tags & 100 tags\\ 
		BeerAdvocate \cite{McAuley:2012:LAA:2471881.2472547} & 33,387 & 1,586,259 beer ratings & 3/91 ratings & 0/3 beer types & 45 types\\
		
		\hline
	\end{tabular}
 	\caption{A summary of datasets used in this paper. $|V|$ reports the total number of nodes (users), $|A|$ the total count of non-zero attribute values (e.g. plays, ratings). We report the median and 90th percentile of unique node attributes (e.g. the number of unique artists a user listens to) all nodes. We also report the median and 90th percentile count of non-zero labels per node (e.g. how many genres `this' user is a listener of).} 
 	\label{tab:data}
 \end{table*}

We demonstrate our methodology on three datasets of user activity data. This data includes beer review history from BeerAdvocate, music listening history from Last.fm, and movie rating history from MovieLens.

\subsection{BeerAdvocate}

BeerAdvocate is a website founded in 1996 hosting user-provided text reviews and numeric ratings of individual beers. The BeerAdvocate review dataset \cite{McAuley:2012:LAA:2471881.2472547} contains 1.5M beer reviews of 264K unique beers from 33K users. We summarize each review by the average rating across 5 review categories on a 0-5 scale (`appearance', `aroma', `palate', `taste', `overall'), yielding node attribute vectors where non-zero elements are the user's average rating for `this' beer (Details in Table \ref{tab:data}).

\subsubsection{Genre Labels} 

The BeerAdvocate dataset contains a field of beer `style,' with 104 unique values. We consider a user a `reviewer' of a particular beer style if they've reviewed at least 5 beers of the style. We further prune styles with fewer than 100 reviewers to yield 45 styles (e.g. `Russian Imperial Stout', `Hefeweizen', `Doppelbock'). A labelset $L(`style') : l_i \in \{0, 1\}$ is generated by the indicator function for whether user $v_i$ is a reviewer of `style.' 

\subsection{Last.fm}

The Last.fm social network is a platform focused on music logging,recommendation, and discussion. The platform was founded in 2002, presenting many opportunities for longitudinal study of user preferences in social networks.

The largest connected component of the Last.fm social network and all associated music listening logs were collected through March 2016. Users in this dataset are ordered by their discovery in a breadth-first search on the explicit `friendship' network from the seed node of a user account opened in 2006. Previous authors sample a dataset of the first $\approx 20K$ users, yielding a connected component with 678K edges, a median degree of 35, and over 1B plays collectively by the users over 2.8M unique artists. For each node, sparse non-zero attribute values correspond to that user's total plays of the unique artist. The Last.fm `explicit' social network of declared `friendship' edges is given as input to our framework as one of several network models. Other network models are inferred on attributes.

\subsubsection{Genre Labels} 

We use the last.fm API to collect crowd-sourced artist-genre tag data. We describe a user as a `listener' of an artist if they've listened to the artist for a total of at least 5 plays. A user is a `listener' of a genre if they are a listener of at least 5 artists in the top-1000 most tagged artists with that genre tag. We generate the labelset for `genre' as an indicator function for whether user $v_i$ is a listener of this genre. We collect artist-genre information for 60 genres, which are hand-verified for artist ranking quality.

\subsection{MovieLens}

The MovieLens project is a movie recommendation engine created in 1995. The MovieLens 20M ratings dataset \cite{Harper:2015:MDH:2866565.2827872} contains movie ratings (1-5 stars) data over $138,493$ users, through March 2015. Non-zero values in a node's attribute vector correspond to the user's star rating of `this' unique film.

\subsubsection{Genre and Tag Labels}

We generate two different types of labelsets on this data. Each movie includes a coarse `genre' field, with 19 possible values (e.g. `Drama', `Musical', `Horror'). For each genre value, we generate a binary node label-set as an indicator function of whether `genre' is this node's  highest-rated genre on average (according to star ratings). 

This collection of labelsets is limiting because each node has a positive label in exactly one labelset. This means that we cannot build node-level statistics on task performance over multiple positive instances. 

To address this, we also generate labels from user-generated tags. We find the top-100 tags, and the top-100 movies with the highest frequency of the tag. A user is a `viewer' of `this' tag if they have rated at least 5 of the these top-100 movies. Our binary `tag' labels are an indicator function for this `viewer' relationship. These tags include more abstract groupings (e.g. `boring',`inspirational', `based on a book', `imdb top 250') as well as well-defined genres (e.g. `zombies', `superhero', `classic'). 

\subsection{Validation, Training and Testing Partitions}
We split our three datasets into contiguous time segments for validation, training, and testing, in this order. This choice is so models are always validated or tested against adjacent partitions, and testing simulates `future' data. These partitions are roughly 2-7 years depending on the dataset, to produce roughly equal-frequency attribute partitions. For all of our label definitions, we evaluate them per partition, so a user could be a reviewer/listener/viewer in one partition and not another, according to activity.

For the Last.fm explicit social network, we do not have edge creation time between two users. For the LP task, edges are split at random $50\%$ to training, and $25\%$ to validation and testing. We sample non-edges disjoint from the union of all time segments, e.g. sampled non-edges appear in none of training, validation or test edge-sets. This ensures a non-edge in each partition is a non-edge in the joined network. 




\subsection{Interpreting Tasks on Our Datasets}

On each of these datasets, what do the link prediction (LP) and collective classification (CC) tasks measure for each network model? For CC, we construct `listener', `viewer', and `reviewer' relationships, which are hidden function mappings between sets of items (unique artists, movies, beers) and tags. The network model is evaluated on how well it implements all of these functions under a unified relational model (queried by a task at some locality). We then evaluate how well this learned function performs over time.  

The LP task measures the stability of underlying similarity measures over time. On datasets such as Last.fm and MovieLens, new artists/films and genres emerge over the time of data collection. LP measures whether learned discriminative artists/movies at some locality predict the similarity ranking (e.g. the ultimate edge/nonedge status) of future or past node attribute distributions, while the constituent contents of those distributions may change.








\section{Evaluation}

We validate network models under varying configurations (similarity measure, network density, task locality and task method) on each dataset, over the dataset's defined $L^*$ set of labelsets. We measure \textit{precision} on both collective classification (CC) and link prediction (LP). Over all network model configurations, we rank models on precision evaluated on the validation partition, and \textit{select} the top ranked model to be used in testing. To evaluate robustness of model selection, we evaluate all network model configurations on both validation and testing partitions to closely examine their full ranking.

Let $p_i$ denote the precision of the $i$-th network model configuration, $p_s$ as the precision of the selected model configuration evaluated on the testing partition,  $p_{(1)}$ as the precision of the `best' model under the current context, and more generally $p_{(10)}$ as the vector of the top-10 precision values.
 
\subsection{Model Stability: Precision}
\label{subsec:performance}

\begin{table}
	\centering
	\scalebox{\boxscale}{
	\begin{tabular}{|cccc|ccc|}
\hline
\multicolumn{4}{|c|}{Precision (Testing)} & \multicolumn{3}{c|}{Validation vs. Testing} \\
\hline
\multicolumn{1}{|c}{Task-Method}   & $\mu$ & $\mu_{(\topk)}$  & $p_{(1)}$ & $\Delta \mu$ & $\Delta \mu_{(\topk)}$ & $\Delta p_{(1)}$ \\
\hline \hline
\multicolumn{7}{|c|}{BeerAdvocate} \\
 \hline

CC-RF & 0.12 & 0.20 & 0.23 & 0.07 &	0.12 & \textbf{-0.01}\\
CC-SVM & 0.35 &	0.64 & 0.70 & -0.06	& -0.08 & \textbf{-0.03} \\
\rowcolor{Gray}
LP-RF & 0.50 &	0.53& 0.58 & 0.03	&      0.06 & -0.11 \\
\rowcolor{Gray}
LP-SVM &0.51 &	0.57& 0.64 & 0.03	&   0.06 & -0.04 \\

\hline

\multicolumn{7}{|c|}{Last.fm} \\
 \hline

       CC-RF &  0.18 &       0.38 & 0.39    &    0.04 &              0.08 &\textbf{-0.01} \\
      CC-SVM &  0.38 &       0.62 &  0.64   &    0.03 &              0.06 &\textbf{-0.01}\\
\rowcolor{Gray}
      LP-SVM &  0.53 &       0.60 & 0.68    &   -0.01 &             -0.01 &\textbf{-0.00}\\
\hline

\multicolumn{7}{|c|}{MovieLens: Genres} \\
 \hline

       CC-RF &  0.01 &       0.03 & 0.06    &    0.02 &              0.02 &-0.05\\
      CC-SVM &  0.15 &       0.30 & 0.36    &   -0.03 &             -0.03 &\textbf{-0.01}\\
\rowcolor{Gray}
       LP-RF &  0.46 &       0.48 & 0.6    &    0.06 &              0.07 &-0.21\\
\rowcolor{Gray}
      LP-SVM &  0.45 &       0.47 &  0.52   &    0.08 &              0.08 &-0.09\\
\hline
\multicolumn{7}{|c|}{MovieLens: Tags} \\
 \hline
CC-RF & 0.28 & 0.60 & 0.68 & 0.22 & 0.17 & -0.10 \\
CC-SVM & 0.55 &	0.80 & 0.86& 0.04	& 0.07 &-0.06\\
\hline
\end{tabular}

}
	\caption{Task precision over all datasets and methods. $\mu$ reports the mean precision over all network models configurations. $\mu_{(\topk{})}$ reports the mean precision over the top \topk{} ranked configurations. $\Delta\mu$ reports the mean of the differences in precision for a given network configuration, between validation and testing partitions (0 is best). $p_{(1)}$ reports the precision of the top-ranked model configuration. $\Delta p_{(1)}= p_{s} - p_{(1)} $ reports the precision of the model `$s$' selected in validation, evaluated on the testing partition, minus precision of the best model in testing (0 is best). Link prediction tasks (grey rows) are base-lined by $0.5$, which is random in the balanced prediction setting. \textbf{Bold} indicates $\Delta p_{(1)} \leq 0.1(p_{(1)} - \mu)$. }
	\label{tab:perf}
\end{table}

Table \ref{tab:perf} reports the mean precision over all network configurations evaluated on the testing partition. A data point in this distribution is the precision value of one network configuration, organized by each row's task method. $\mu$ reports the mean precision over \textit{all} such configurations ($|N|> 100$). This represents a baseline precision from any considered modeling choice. $\mu_{(\topk{})}$ reports the mean precision over the top-\topk{} ranked configurations. $p_{(1)}$ reports the precision of the best network model configuration for `this' task method. The difference between $p_{(1)}$ and $\mu$ represents the maximum possible gain in precision added by model selection. 

Table \ref{tab:perf} reports $\Delta \mu$, the stability of mean precision over validation and testing partitions. A data point in this distribution is $p_{i, \mathrm{validation}} - p_{i, \mathrm{testing}}$ the difference in precision for the same `$i$' model configuration in validation and testing partitions. Positive values are more common in Table \ref{tab:perf}, indicating better aggregate performance in the validation partition. This matches our intuition that relationships among new items found in the testing partition may be more difficult to learn than preceding validation data, which is largely a subset of items found in the training partition. 
 
The best model in validation need not be the best possible model in testing, especially with many similar models. Table \ref{tab:perf} reports $\Delta p_{(1)}= p_{s} - p_{(1)}$, the difference in precision between the selected model configuration, and the best possible model configuration, both evaluated on the testing partition (0 is best). We highlight values in bold with $\Delta p_{(1)} \leq 0.1(p_{(1)} - \mu)$, i.e. less than $10\%$ of the possible lift in the testing evaluation. $\Delta p_{(1)}$ is one of many possible measures of network model robustness. We look more closely at the selected model, as well as stability in deeper model rankings.

\subsection{Model Consistency: Selected Model Ranking}

\addtolength{\tabcolsep}{-3pt}   
\begin{table}[ht]
	\centering
	\scalebox{\boxscale}{

\begin{tabular}{|cc|cc|}
\hline
\multicolumn{4}{|c|}{Selected Model-Locality and $\mathrm{rank}$} \\
\hline
 CC-RF & CC-SVM & LP-RF & LP-SVM \\
\hline \hline
\multicolumn{4}{|c|}{BeerAdvocate}\\
\hline
\textbf{0.99} & \textbf{0.99} & 0.09 & \textbf{0.99} \\
\textbf{KNN-Local}& \textbf{TH-Local} & KNN-Ensemble & TH-Ensemble\\
\hline
\multicolumn{4}{|c|}{Last.fm}\\
\hline
\textbf{0.90} & \textbf{0.91} & -- & \textbf{1.00} \\
\textbf{KNN-Community} &    \textbf{Social-Community} &   --&        \textbf{TH-Local} \\
\hline
\multicolumn{4}{|c|}{MovieLens: Genres}\\
\hline
0.61 & \textbf{0.95} & 0.06 & 0.29 \\
KNN-Local &           \textbf{TH-Community} &           TH-Ensemble & TH-Ensemble \\
\hline
\multicolumn{4}{|c|}{MovieLens: Tags}\\
\hline
\textbf{0.92} & 0.87 & -- & -- \\
KNN-Local &         KNN-Local &    -- & -- \\
\hline

\end{tabular}
	}
	\label{tab:results_rank}
	\caption{The normalized rank of the selected model, evaluated in the testing partition. $\mathrm{rank}=1$ indicates the best models in both validation and testing partitions are the same. The row below the $\mathrm{rank}$ indicates the network model and locality selected in validation. \textbf{Bold} indicates all selected models with $\mathrm{rank} \geq 0.9$, i.e. in the top $10\%$ of model configurations.}
	\label{tab:rank_diff}
\end{table}
\addtolength{\tabcolsep}{3pt}   

Table \ref{tab:rank_diff} reports the normalized $\mathrm{rank}$ of the selected model configuration, evaluated on the testing partition. Values can be interpreted as percentiles, where 1 indicates the same model is top-ranked in both partitions (i.e. higher is better). We highlight models with high stability in precision between validation and testing partitions: $\mathrm{rank} > 0.9$, i.e. the selected model is in the top $10\%$ of model configurations on the testing partition.

Table \ref{tab:rank_diff} shows several cases of rank inconsistency for particular problem settings (e.g. BeerAdvocate LP-RF, MovieLens LP-RF) and notable consistency for others (BeerAdvocate CC-RF, CC-SVM) ranked over many total network configurations ($|N| > 100$). This is a key result demonstrating that appropriate network models change according to the task and the underlying dataset. For Last.fm, community localities are selected for both CC methods. The social network and community locality are selected for CC-SVM. This is very surprising from previous results show poor performance of local models on the social network but did not evaluate other localities \cite{mlg2017_13}. For CC on BeerAdvocate, local models are consistently selected and have a high rank in testing, even though SVM and RF methods have very different performance in absolute precision. This might indicate that preferences are more local in the BeerAdvocate population than Last.fm. In this way, interpretability of locality, network model, and underlying similarity measures can drive further hypothesis generation and testing, especially for network comparison across domains.

\subsection{Model Stability: Rank Order}

\begin{table}[htbp]
	\centering
	\scalebox{\boxscale}{
	\begin{tabular}{|c|ccc||c|}
\hline
\multicolumn{5}{|c|}{Rank Ordering (Validation vs. Test)} \\
\hline
 \multicolumn{1}{|c}{Task-Method}  & $\tau$ & $p$-value &  $\mathrm{intersection}_{(10)} $ & \textbf{Total} \\
\hline \hline
\multicolumn{5}{|c|}{BeerAdvocate}\\
\hline
CC-RF & 0.7   &\textbf{1.75E-34} & \textbf{6} & 4 \\
CC-SVM & 0.6  &\textbf{1.54E-25} & \textbf{8} & 4 \\
LP-RF & 0.34  & \textbf{3.08E-09} & 3  & 1 \\
LP-SVM & 0.44 &\textbf{1.09E-14} & \textbf{5} & 3  \\
\hline
\multicolumn{5}{|c|}{Last.fm}\\
\hline
CC-RF &   0.88 & \textbf{2.35E-24} & \textbf{9} & 4 \\
CC-SVM &  0.88 & \textbf{5.85E-21} & \textbf{8} & 4 \\
LP-SVM &   0.70& \textbf{8.15E-14} & \textbf{8} & 4 \\ 
\hline
\multicolumn{5}{|c|}{MovieLens: Genres}\\
\hline
CC-RF &   0.15 & 6.89E-03          & 0 & 0\\
CC-SVM &  0.57 & \textbf{9.99E-24} & 4 & 3\\
LP-RF &  -0.07 & 2.29E-01          & 0 & 0\\
LP-SVM & -0.07 & 2.40E-01          & 0 & 0\\
\hline
\multicolumn{5}{|c|}{MovieLens: Tags}\\
\hline
CC-RF & 0.61 & \textbf{8.95E-27} & 0 & 2\\
CC-SVM &0.52 & \textbf{4.43E-20} & 1 & 1\\

\hline
\end{tabular}

}
	\label{tab:results_full}
	\caption{The Kendall's $\tau$ rank order correlation between models in validation and test partitions, according to precision. 1 indicates the rankings are the same, 0 indicates random ranking. $\tau_{\topk{}}$ reports rank order correlation on the top-\topk{} models. We report associated p-values. \textbf{Bold} indicates the models with $p < 1.00\mathrm{E}{-03}$ and $\mathrm{intersection}_{10} \geq 5$.}
	\label{tab:tau}
\end{table}

Table \ref{tab:tau} reports the Kendall's $\tau$ rank order statistic between the ranking of model configurations by precision, for validation and testing partitions where $\tau=1$ indicates the rankings are the same. We report the associated $p$-value of the $\tau$ rank order statistic. For several tasks on several data-sets, ranking is very consistent over all model configurations. 

While this ranking shows remarkable consistency, it's not suitable when the result contains many bad models, which may dominate $\tau$ at low ranks. Due to this, $\mathrm{intersection}_{(10)}$ reports the shared model configurations in the top-10 of validation and testing partitions. Since top-$k$ lists may be short and have disjoint elements, we find the simple intersection rather than rank order. We highlight tasks in bold at a rank order significance level of $p < 1.00\mathrm{E}{-03}$, and $\mathrm{intersection}_{(10)} \geq 5$.

Table \ref{tab:tau} `Total' summarize the count of bold entries across Tables \ref{tab:perf}, \ref{tab:rank_diff}, and \ref{tab:tau}. This corresponds to scoring the network model on (1) precision stability, (2) selected model rank consistency, (3) full ranking stability, and (4) top-$10$ ranking consistency.

MovieLens under `tag' labels is a peculiar result. It performs very well at both $\mu_{(10)}$ and $p_{(1)}$ for both SVM and RF. However it has a high $\Delta p_{(1)}$ and low $\mathrm{intersection}_{(10)}$. Looking closer at the results, two similar groups of local model perform well. However, in validation, this is under `adjacency' locality, and the testing partition favors the wider `BFS' local configurations. One challenge to address is appropriately grouping models where the ranking of specific configurations is uninformative and can introduce ranking noise, but the ranking between categories (e.g. locality) is informative. 

MovieLens improves learning by several factors by using `tag' labelsets where each node may have several positive instances rather than a single positive instance. BeerAdvocate and Last.fm have very clear signals of robust selected models over our scoring criteria. 

\subsection{Consistency: Task Method Locality}

\begin{figure}[htbp]
\centering
\includegraphics[width=\figwidth\columnwidth]{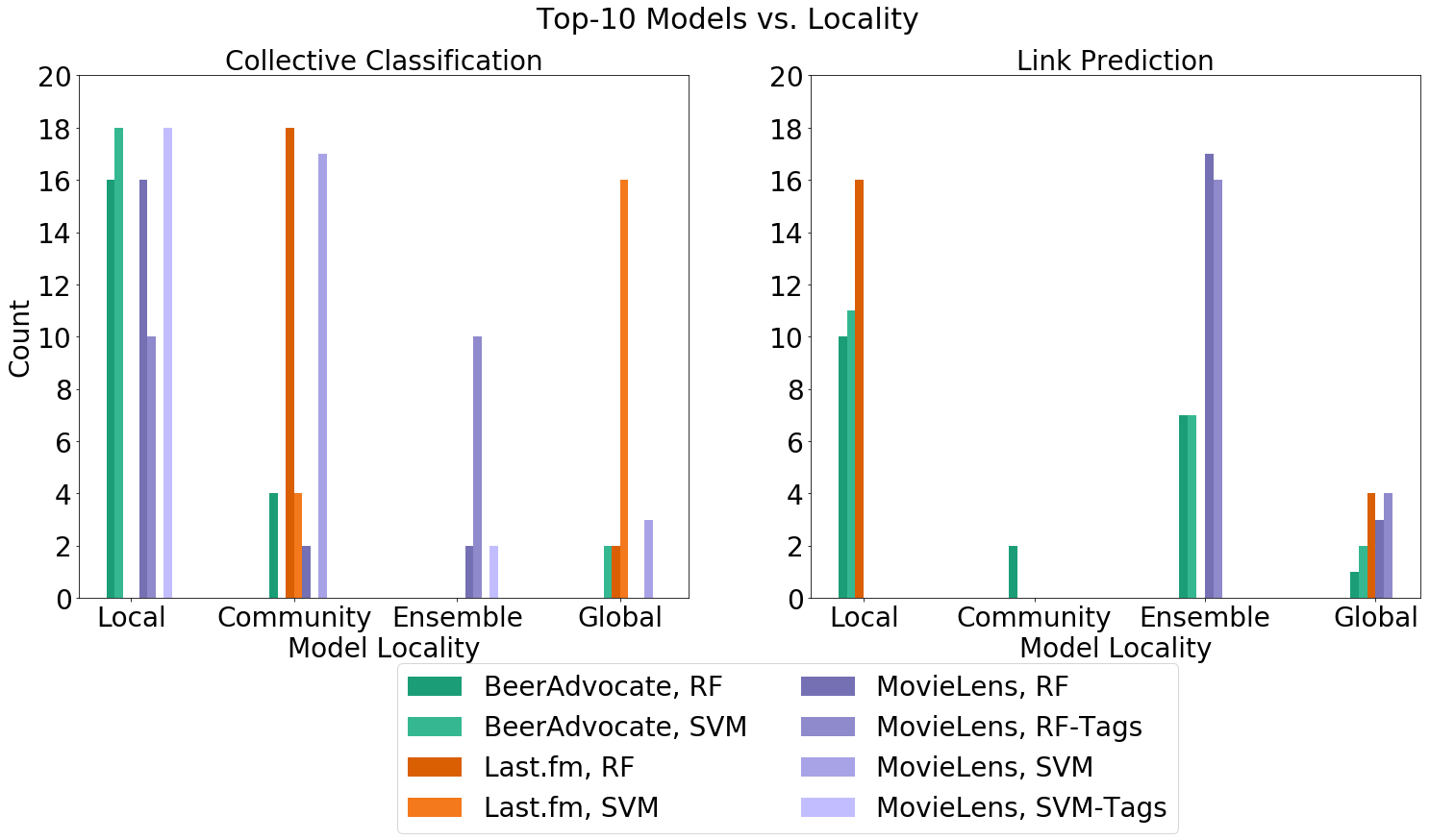}
\caption{Counts of task locality associated with the top-\topk{} ranked models in validation and testing partitions (\twotopk{} total). Primary colors denote different datasets, shades denote different task methods.}
\label{fig:locality}
\end{figure}

Our framework allows further investigation of localities suitable for particular types of tasks, measured by their ranking. Figure \ref{fig:locality} reports the counts of model configurations at varying localities for the top-\topk{} model configurations in the validation and testing partition (\twotopk{} total), for CC (left) and LP (right). Each principal color represents a dataset, and shades denote different task methods. 

Collective classification on BeerAdvocate strongly favors local task localities, and Last.fm favors community and global localities; both of these agree with model selections in Table \ref{tab:rank_diff}. `Global' locality measures the extent that population-level models are favored to any locality using network information. Looking closer at Last.fm, the $\Delta p_{(1)}$ for the best-ranked `Global' configuration in testing is only -0.01 for CC-SVM, and -0.05 for CC-RF. This indicates a very weak network effect on Last.fm for CC under our explored models.

From CC to LP tasks (left to right), model ranking preferences change greatly per dataset. BeerAdvocate increases preference for ensemble methods (primarily ``Sum of Attributes'', see Section \ref{subsec:task_locality}). The preference for global locality largely disappears on Last.fm for LP, instead the task has a very strong preference for local models (all using `adjacency,' rather than BFS local models). This demonstrates that we find robust indicators for models and localities suited for different tasks, which change both by dataset and task.

\begin{figure}[htbp]
\centering
\includegraphics[width=\figwidth\columnwidth]{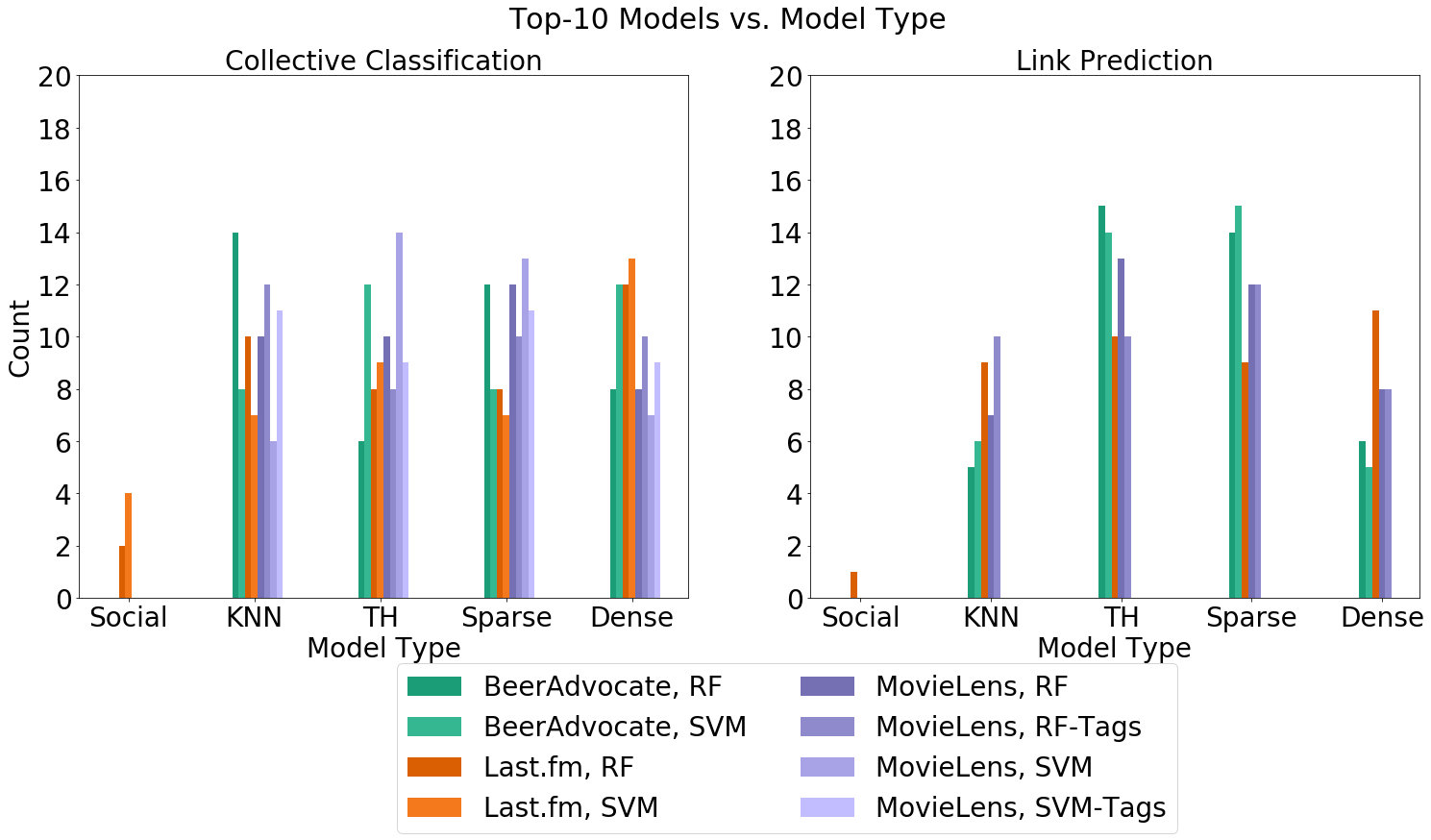}
\caption{Counts of network models and density associated with the top-\topk{} ranked models in validation and testing partitions (\twotopk{} total). Primary colors denote different datasets, shades denote different task methods.}
\label{fig:modeltype}
\end{figure}

Figure \ref{fig:modeltype} reports similar counts across types of network models and densities for the top-\topk{} model configurations. The first three bar groups report a total of \twotopk{} model configurations over `Social' (only Last.fm), `KNN,' and `TH' network models. The next two bar groups report \twotopk{} configurations over `Sparse' and `Dense' settings. `Sparse' refers to \textit{very} sparse networks on the order of $0.0025$ density, while `dense' is on the order of densities ordinarily observed in social networks (e.g. $0.01$). In the case of the Last.fm social network, factors on observed density $[0.75, 1]$, (e.g. $0.75\times \mathrm{d}(E)$) are considered dense and $[0.25, 0.5]$ are sparse.

For all tested datasets, there is not a strong preference for a particular network model or density for either CC or LP. However, this does not mean that precision is `random' over varying network models, or that other types of underlying data may have model preferences for our set of models. The $\tau$ rank order and $\mathrm{intersection}_{10}$ are very consistent in several of these task instances (Table \ref{tab:tau}). Instead, locality preferences seem to drive the three datasets we examine, where network models will perform similarly under the same locality than across localities. 

\begin{table}
	\centering
	\scalebox{\boxscale}{
	\begin{tabular}{|c|cc|cc|ccc|}
\hline
\multicolumn{8}{|c|}{Median $p_i - p_j$, match vs. mismatch (SVM)} \\
 \multicolumn{1}{|c}{} &  \multicolumn{2}{c|}{BeerAdvocate}  & \multicolumn{2}{c|}{Last.fm} & \multicolumn{3}{c|}{MovieLens} \\
\multicolumn{1}{|c}{} & \multicolumn{1}{c}{CC}  & LP & CC & LP & CC & CC-Tags & LP \\
\hline \hline

Locality & -0.03 & 0.01 & -0.14 & -0.04 & -0.04 & 0.01 & -0.02 \\
Model & 0.00 & 0.00 & 0.01 & 0.01 & 0.00 & 0.00 & 0.00 \\
\hline
\end{tabular}
}
	\caption{The difference between medians of pairwise network model configurations grouped by matching or mismatching criteria. More negative values denote higher median precision difference among mismatching configuration pairs.}
	\label{tab:matchmismatch}
\end{table}

We evaluate this hypothesis in Table \ref{tab:matchmismatch}. We report the median of pairwise differences of precision between configuration pairs by matched and mismatched models or localities: $\mathrm{median}(p_{i} - p_{j}) - \mathrm{median}(p_{k} - p_{l})$, where $(i,j)$ are all matched pairs grouped by the same locality or model, and $(k,l)$ all mismatched pairs. More negative values represent higher differences in precision on mismatches than matches, for that row's criteria. Mismatching localities indeed account for more difference in precision than mismatching network models.

\subsection{Model Selection and Cross-Task Performance}

\begin{table}
	\centering
	\scalebox{\boxscale}{
	\begin{tabular}{|c|cc|cc|}
\hline
\multicolumn{5}{|c|}{Cross-Task Model Ranking (SVM)} \\
\hline
 \multirow{2}{*}{Model Selection \textbackslash~Testing}  & \multicolumn{2}{c|}{CC-SVM} & \multicolumn{2}{c|}{LP-SVM} \\
 &\multicolumn{1}{c}{$\Delta p_{(1)}$}&$\mathrm{rank}$ & \multicolumn{1}{c}{$\Delta p_{(1)}$}&$\mathrm{rank}$\\
\hline \hline

BeerAdvocate, CC-SVM & -0.03&	0.99&	-0.17 &0.14\\
Last.fm, CC-SVM&-0.01 &	0.91&-0.16&0.68\\
MovieLens, CC-SVM& -0.01	&0.95 &	-0.08 &0.47\\
\hline
BeerAdvocate, LP-SVM & -0.65&	0.09 &-0.04 &	0.99 \\
Last.fm, LP-SVM	&-0.43& 0.21 &-0.00&1.00 \\
MovieLens, LP-SVM &-0.31&0.76 &-0.09&	0.29\\
\hline \hline
\multicolumn{5}{|c|}{Average-Precision Model Selection (SVM)} \\
\hline 
BeerAdvocate, SVM & -0.29&	0.60 &-0.10 &	0.84 \\
Last.fm, SVM	&-0.01& 0.94 &-0.13&0.75 \\
MovieLens, SVM &-0.01&0.98 &-0.09&	0.31\\
\hline

\end{tabular}

}
	\caption{(Upper) Performance of network models selected in the validation partition (left), evaluated on varying tasks (top) according to the difference against the best model configuration in testing $\Delta p_{(1)}= p_{s}-p_{(1)}$, for `$s$' the selected configuration (0 is best), and $\mathrm{rank}$, the normalized rank of the selected model configuration in the testing partition (higher is better). Diagonal entries correspond to values in Tables \ref{tab:perf}, and \ref{tab:rank_diff} for $\Delta p_{(1)}$ and  $\mathrm{rank}$, respectively. (Lower) Average-Precision Model Selection using ranking from the average of precision on CC and LP.}
	\label{tab:cross}
\end{table}

Table \ref{tab:cross} (Upper) reports model performance \textit{across} tasks. We do model selection on the validation partition (for each task on the left) and report task performance in testing, on the task method given by the column. The $\Delta p_{(1)}$  and $\mathrm{rank}$ are calculated as previously, where values on the diagonal are the same as in Tables \ref{tab:perf} and \ref{tab:rank_diff}, respectively. On the off-diagonal, the model is evaluated in testing on the task it was not selected on. Table \ref{tab:cross} (Lower) reports model performance by doing model selection on the average of CC and LP precision.

This result clearly demonstrates the main take-away of our study: the `best' network model depends on the subsequent task. The off-diagonal shows that in every case, models selected on the `other' task perform very poorly. Consider the worst case for same-task selection--LP on MovieLens--scored 0 on our 4 selection criteria (Table \ref{tab:tau}), yet has a selected model performing 3x better in $\Delta p_{(1)}$ than it's cross-task selected model. Over our three datasets, the \textit{average} factor increase in $\Delta p_{(1)}$ performance from model selection using same-task compared to using cross-task is $\approx 10x$. 

Average-Precision Model Selection performs poorly in \textit{both} tasks for BeerAdvocate, and is dominated by the CC task in Last.fm and MovieLens, closely matching the CC rows. Therefore, by selecting a network model on both tasks, we never recover the suitable model for link prediction in any of the three datasets. 

\subsection{Node Difficulty}
\label{subsec:node_difficulty}
 
Both of our prediction tasks evaluate the same node over many predictions. For collective classification, we make a prediction at a node for each positive label instance over many labelsets. For link prediction, we associate predictions on the ego-net (on the order of hundreds), with that node.

\begin{figure}[htbp]
\centering
\includegraphics[width=\subfigwidth\columnwidth]{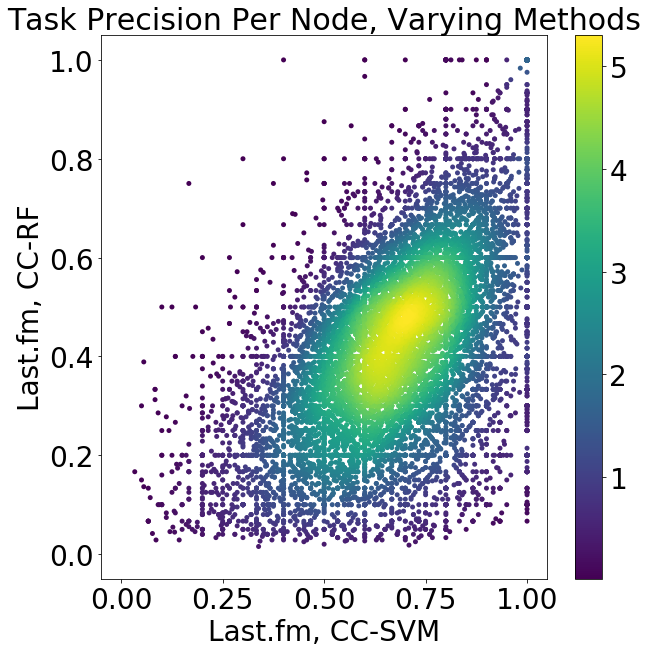}
\includegraphics[width=\subfigwidth\columnwidth]{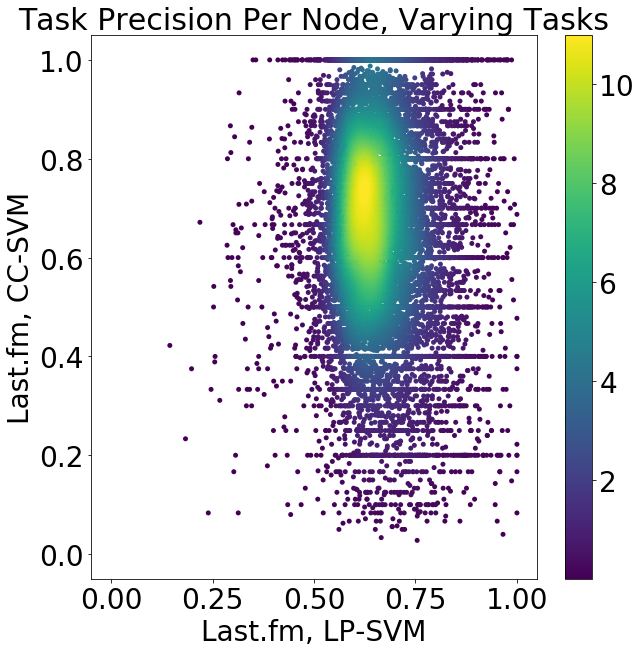}
\caption{The precision per node on Last.fm, aggregated over predictions of the top top-$5$ models in validation and testing partitions (10 models total). (Left) the distribution of nodes, varying task method, (Right) the distribution of nodes, varying task. x and y axes indicate precision over the node's predictions. Color corresponds to the kernel density coefficient.}
\label{fig:heatmap_lfm}
\end{figure}

We can therefore build robust distributions of precision for individual nodes over a particular model configuration or union of configurations. Figure \ref{fig:heatmap_lfm} reports the density map of precision of individual nodes for Last.fm, aggregated over the top-5 models in validation and testing partitions (10 models total), on varying methods (left) and tasks (right). The left plot shows that RF and SVM methods are mostly linearly correlated, with SVM performing better. The right plot shows low variance in LP-SVM, with some skew to higher-performing LP nodes in purple. CC-SVM shows higher variance, with a higher ceiling (the band at y=1) and many poorly-performing nodes below y=0.4.

\begin{figure}[htbp]
\centering
\includegraphics[width=\subfigwidth\columnwidth]{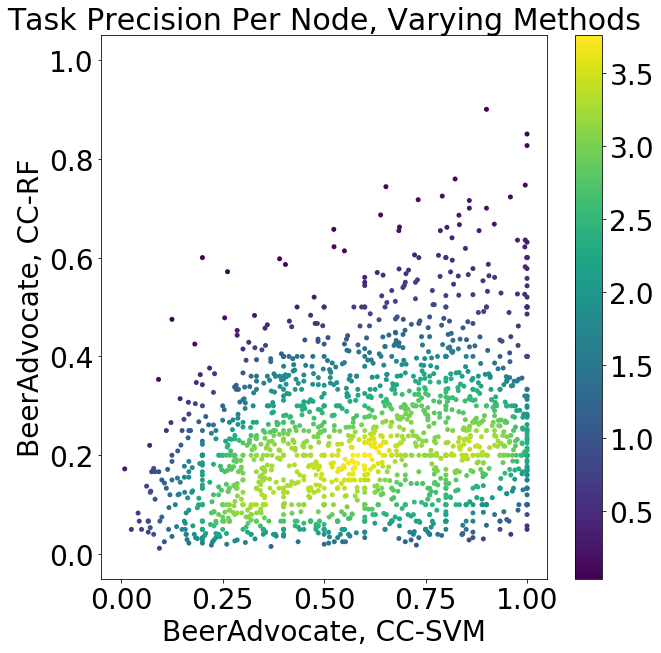}
\includegraphics[width=\subfigwidth\columnwidth]{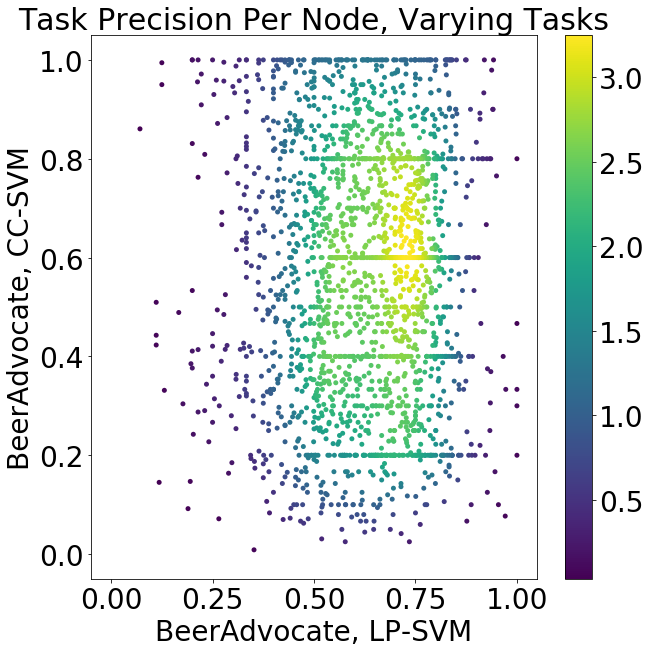}
\caption{The precision per node on BeerAdvocate, as Figure \ref{fig:heatmap_lfm}.}
\label{fig:heatmap_ba}
\end{figure}

Figure \ref{fig:heatmap_ba} reports the same comparison for the BeerAdvocate dataset. Compared to Last.fm, CC-RF performs weaker in relative to CC-SVM, which has higher variance. Comparing CC and LP, the gradient has higher variance in both axes than in Last.fm, with the skew to poorer-performing nodes in LP. 

Measuring the distribution of node hardness over varying multiple  predictions on varying tasks allows further hypothesis generation and testing related to the distribution of these values over the network topology or other feature extraction to characterize the task according to the application. Our focus in this work is primarily on the evaluation methodology for network model selection, so we only give the highest-level introduction of this characterization step.

\section{Conclusion and Future Work}
This work focused on a general task-focused network model selection methodology which uses fundamental network tasks--collective classification and link prediction--to evaluate several common network models and localities. We propose evaluating model selection for network tasks under several criteria including (1) task precision stability, (2) selected model rank consistency, (3) full rank stability, and (4) top-$k$ rank consistency. We evaluate three user rating datasets and show robust selection of particular models for several of the task settings. We demonstrate that network model selection is highly subject to a particular task of interest, showing that model selection across tasks performs an \textit{order of magnitude} better than selecting on another task. 

\subsection{Limitations and Future Improvements}

\subsubsection{Incorporating model cost}
We currently do not incorporate network model cost (e.g. sparsity), nor prediction method cost (e.g. method encoding size in bytes, runtime) as criteria for model selection. In future work we wish to penalize more costly models. 

For example, we train on the order of thousands of small `local' models, while an ensemble model which may have similar performance trains on tens of nodes. Future work will explore ensembles of local methods to \textit{summarize} the task under minimal cost. Some task methods are also fairly robust to our choice of network density parameters; the sparser network model would be preferable.

\subsubsection{Network model Alternativeness}
We would also like to discover \textit{alternative} network models. Model `alternativeness' refers to discovering maximally different model representations (by some criteria) which satisfy given constraints \cite{Qi:2009:PFF:1557019.1557099, niu2010multiple}. In future work we would like to identify maximally orthogonal network models of similar (high) performance over our task and labelset regime, under some informative structural or task orthogonality. Section \ref{subsec:node_difficulty} explores node-level joint density of task performance. Alternativeness in this setting may maximize differences in the joint distributions of well-performing models, and report or merge this set of models in model selection. 

\subsubsection{Model Stationarity}

Our results in Section \ref{subsec:performance} show some indication of improved performance on the preceding partition (validation) than the future partition (testing). Our model selection framework tests network model temporal stationarity `for free,' and can be used to measure the decay of both predictive performance and model rank ordering over increased time-horizons. Both of these signals can indicate a model change in the underlying data over time.

\bibliographystyle{IEEEtran}
\bibliography{acmsmall-sample-bibfile}

\end{document}